\documentclass{iau}

\usepackage{graphicx}
\usepackage[usenames, dvipsnames]{color}

\usepackage{natbib}
\usepackage{txfonts}
\usepackage{url}

\usepackage{hyperref}
 \hypersetup{
     final=true,
     pageanchor=true,
     colorlinks=true,
     breaklinks=true,
     linkcolor=blue,
     citecolor=blue,
     urlcolor=blue,
     pdfpagemode=UseNone,
     pdftitle={Formation and evolution of an active region filament},
     pdfauthor={C. Kuckein, R. Centeno and V. Martinez Pillet},
     pdfsubject={Solar Physics},
     pdfkeywords={Sun: filaments, prominences, Sun: photosphere, Sun: chromosphere, Sun: magnetic
 fields, techniques: polarimetric}}

\newcommand*\degr{\ensuremath{^\circ}}

\newcommand*\arcsec{\ensuremath{^{\prime\prime}}}

\title[Formation and evolution of an active region filament] %% give here short title %%
{Formation and evolution of an active region filament}

\author[C. Kuckein, R. Centeno and V. Mart\'inez Pillet]   
{Christoph Kuckein$^{1,2}$, Rebeca Centeno$^3$ and Valent\'in Mart\'inez Pillet$^{4}$}
\pubyear{2013}
\volume{300}
\pagerange{1--2}
\jname{Nature of Prominences and their Role in Space Weather}
\editors{B.\ Schmieder, J.-M.\ Malherbe \& S.\ Wu, eds.}

\affiliation{$^1$Leibniz-Institut f\"ur Astrophysik Potsdam (AIP), \\
An der Sternwarte 16, 14482, Potsdam, Germany.  \\
email: {\tt ckuckein@aip.de} \\
[\affilskip]
$^2$Instituto de Astrof\'\i sica de Canarias (IAC), V\'\i a
L\'{a}ctea s/n, 38205, La Laguna, Tenerife,
Spain \\ [\affilskip]
$^3$High Altitude Observatory (NCAR), Boulder, CO 80301, USA \\ [\affilskip]
$^4$National Solar Observatory (NSO), Sunspot, NM 88349, USA \\}

\pubyear{2013}
\volume{300}
\pagerange{1--2}
\jname{Nature of Prominences and their Role in Space Weather}
\editors{B.\ Schmieder, J.-M.\ Malherbe \& S.\ Wu, eds.}

\begin{document}

\maketitle

\begin{abstract}
Several scenarios explaining how filaments are formed can be found in literature. In this paper, we analyzed the
observations of an active
region filament and critically evaluated the observed properties in the context of current filament formation models.
This study is based on multi-height spectropolarimetric observations. The inferred vector magnetic field has been
extrapolated starting either from the photosphere or from the chromosphere. The
line-of-sight motions of the filament, which was
located near disk center, have been analyzed inferring the Doppler velocities. We conclude that a part of the magnetic
structure emerged from below the photosphere. 

\end{abstract}

\firstsection % if your document starts with a section,
\section{Introduction}
Owing to new observations and a continuous improvement of simulations, in the last years the solar community has
published several works concerning the formation and evolution of filaments. Active region (AR) filament studies are
still scarce in literature, and it is not clear how similar their formation process is to that of the quiescent (QS)
ones. Generally speaking, there are two proposed scenarios that try to explain the formation of filaments: (1) the
sheared arcade (SA) model and (2) the flux rope emergence (FRE) model. The difference between both models mainly 
exists in the atmospheric height of filament formation. On the one hand, in the SA model the filament is formed in the
corona by shearing motions in addition to converging flows at the polarity inversion line (PIL) and reconnection
processes. As a result, the magnetic structure can be a flux rope \citep[e.g.,][]{vanballe89}
or a dipped arcade \citep[e.g.,][]{antiochos94}. On the other hand, the FRE model assumes that a
flux rope emerges from below the photosphere ascending into the corona \citep[e.g.,][]{okamoto08}.
\citet[][]{kuck2} provide an extensive discussion and references related to these two models. In
this work, we present a multi-height study of the magnetic structure of an AR filament observed in July 2005 and
investigate whether this filament fits into the SA or FRE model.

\section{Observations}
The present analysis is mainly based on spectropolarimetric observations acquired with the Tenerife Infrared
Polarimeter attached to the Vacuum Tower Telescope (VTT) in Tenerife \citep[TIP-II;][]{tip2}. The
data sets of an AR
filament belonging to NOAA 10781 were taken close to disk center ($\mu \sim$0.91 and 0.95) on 2005 July 3 and 5.
However, H$\alpha$ images from Big Bear Solar Observatory \citep[BBSO;][]{denker99} show
that the filament was already present a few days
before, on July 1. Line-of-sight (LOS) magnetograms from the Michelson Doppler Imager \citep[MDI,][]{mdi} show
that below the filament an extensive facular region is seen with two polarities clearly separated by
the PIL. Between July 1--7 the opposite polarities became more compact,
almost touching each other at the PIL, and then broadened again. During this event, with characteristics of the
``sliding
door'' effect  \citep[an effect firstly described by][]{okamoto08}, new pores and orphan penumbrae
emerged at the PIL \citep[see][for a detailed description and images]{kuck1,kuck2}.
\citet[][]{okamoto08} related this effect to the emergence of a flux rope below an AR filament. 
  
\sloppy The spectral region observed with TIP-II comprised the photospheric Si \textsc{i} 10827~\AA\ line, the
chromospheric He
\textsc{i} 10830~\AA\ triplet, and two telluric lines with a spectral sampling of $\sim$11.04~m\AA~px$^{-1}$.
Therefore, with this instrument it is possible to simultaneously analyze the vector magnetic field at two different
heights in the solar atmosphere. In this work, we will concentrate on the second set of spectropolarimetric data taken
on July 5 between 7:36 and 14:51 UT. The observing strategy was to scan the filament with the slit parallel to the PIL
with a scanning step size of 0.3\arcsec. The pixel size along the slit was 0.17\arcsec. 

The vector magnetic field was inferred by
carrying out inversions of the four Stokes parameters with two different inversion codes: (1) for the Si~\textsc{i}
line we used the SIR code \citep[Stokes Inversion based on Response functions;][]{SIR} and (2) for the He~\textsc{i}
triplet we used a Milne-Eddington-based inversion code
\citep[MELANIE;][]{MELANIE}. The 180\degr-ambiguity was solved using the AZAM code \citep[][]{AZAM}. The LOS
velocities were determined from the inversions and converted to an absolute scale, i.e.,
corrections for orbital motions and the gravity shift were made \citep[see][for a full description of the velocity
calibration]{kuck3}.

\section{Results}
The inferred magnetic field strength in this AR filament is 600--800~G \citep[][]{kuck1,kuck2},
which was the strongest value detected so far inside AR filaments. Recently, other authors have also reported similar
strong magnetic fields inside filaments, \citep[e.g.,][]{guo10,xu12}. Therefore,
it seems to be rather common to find these strong fields inside AR filaments.

The set of inferred chromospheric and photospheric vector magnetograms 
indicate that the magnetic structure which
supports the filament is a flux rope. To substantiate this result, non-linear force-free (NLFF) field extrapolations,
starting from the photosphere and chromosphere, were carried out. The extrapolations confirmed the presence of
a flux rope structure, which lay surprisingly low in the atmosphere and had its axis located at $\sim$1.4~Mm
above the solar surface \citep[see][for a complete description of the extrapolations and
results]{yelles12}.

\begin{figure}[t]
% \vspace*{-2.0 cm}
\begin{center}
 \includegraphics[width=\textwidth]{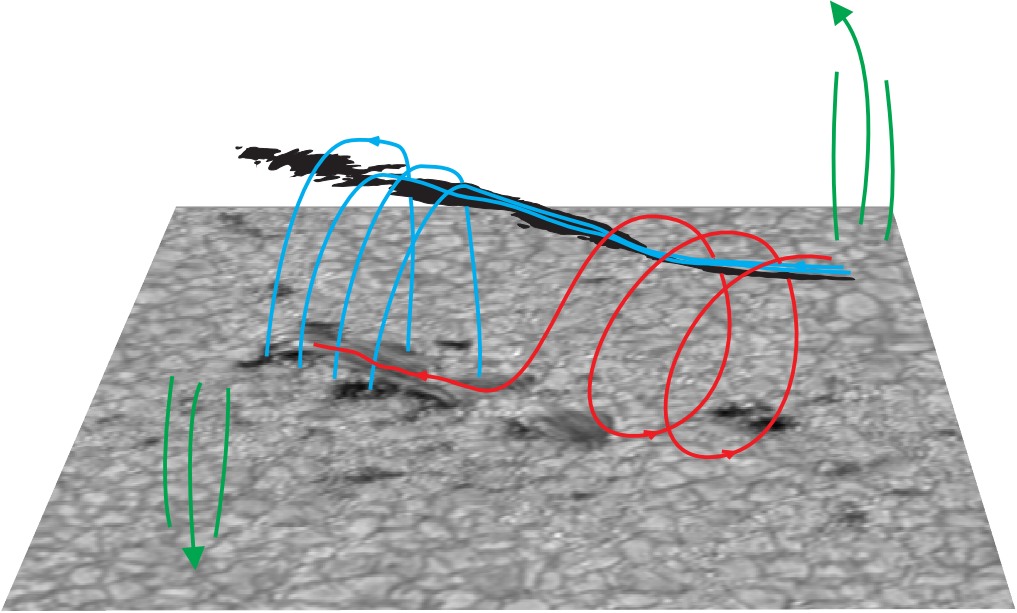} 
% \vspace*{-1.0 cm}
 \caption{Sketch showing the inferred magnetic structure of the AR filament. 
 The surface corresponds to a DOT continuum image. 
 The dark red field lines are representative of what was inferred from the Si \textsc{i} vector magnetic field 
 whereas the light blue lines represent the He \textsc{i} vector magnetic field. 
 The filament is outlined by the black structure taken from the DOT H$\alpha$ image. 
 The light green field lines on both sides of the filament represent the fields in the positive 
 (upward arrow) and  negative (downward arrow) faculae. 
 This figure is from \citet[][]{phd}.}
   \label{fig1}
\end{center}
\end{figure}

Based on the vector magnetograms, we constructed the sketch
presented in Fig.~\ref{fig1}. The cartoon shows a gray-scale continuum image from the Dutch Open Telescope (DOT) on
2005 July 5. In the middle of the image, the aforementioned pores and orphan penumbrae that started to appear on
July 4 are seen. The reconstruction of the filament is represented as a black structure which was extracted from the
corresponding
DOT H$\alpha$ image. The filament follows the PIL. The positive (negative) polarity is at the right (left) side
of the PIL when viewing the figure from the lower right corner. There are three different magnetic field
lines represented in the cartoon. (1) The dark red field lines are based on the photospheric vector
magnetograms. These field lines are aligned along the PIL where pores and orphan penumbrae are seen. However, outside
the
orphan-penumbrae region, the field lines have an inverse polarity configuration (pointing
from negative to positive polarity). (2) The light blue field lines were derived from the chromospheric vector
magnetograms. In
this case, the field lines are parallel to the filament axis outside the orphan-penumbrae region, whereas inside, the
field lines show a normal configuration (the field lines point from positive to negative polarity). (3) The green field
lines, close to the corners of the image, represent the positive (upward pointing arrow) and negative (downward
pointing arrow) polarity of the faculae. 

To shed light on the formation of the flux rope it was crucial to infer the LOS velocities. Calibrations
using two telluric lines and corrections related to orbital motions and the gravity shift were carefully carried out
\citep[see][]{kuck3}. 

The motions of the transverse magnetic fields in the photosphere, in the orphan-penumbrae
area, show, on a seven-hour average, a slow upward trend. Above, in the chromosphere, the filament moves on average
downward. Nevertheless, there are clearly localized upflow areas \citep[see Fig. 3 in][]{kuck3}.
This indicated, when looking at the orphan-penumbrae area in Fig. \ref{fig1}, that the axis of the flux rope is
slowly rising. Half of the flux rope is below the surface, which seems to be responsible for the formation of the
orphan penumbrae. The upper part of the flux rope reaches the chromosphere where groups of field lines produce the
upflow patches detected in the chromosphere. The portion of the filament that does not have orphan penumbrae below
behaves differently. Out of seven maps, the first four show on average upward motions of the transverse fields. The
other three show on average velocities close to zero. The chromospheric counterpart shows downward motions. 

\section{Discussion}
The initial formation phase of this AR filament cannot be described with the presented data sets because the filament
was
already present prior to our observations. However, we have shown that AR filaments can have extremely low-lying flux
rope structures (even as low as the photosphere) that support the filament. This
structure can eventually emerge from below the photosphere, as seen in our data sets, generating pores and orphan
penumbrae along the PIL. Therefore, AR filaments can have a photospheric counterpart. In the chromosphere, the
filament's plasma is being pushed upward by the magnetic field that expands from the emerging flux rope structure. At
the same time, the emerging flux rope supports the filament material against gravity. 
Altogether, the present results favor a flux rope emergence scenario.

\begin{acknowledgements}
\noindent CK greatly acknowledges the travel support received from the IAU. The VTT is operated by the
Kiepenheuer-Institute for Solar Physics in Freiburg, Germany, at the Spanish Observatorio del Teide, Tenerife, Canary
Islands. The National Center for Atmospheric Research (NCAR) is sponsored by the National Science Foundation (NSF). The
authors would like to thank C. Denker for carefully reading the manuscript.
\end{acknowledgements}

\end{document}